\documentclass{article}
\usepackage{amsmath}
\usepackage{graphicx,psfrag,epsf}
\usepackage{enumerate}
\usepackage[utf8]{inputenc}
\usepackage[backend=biber,sorting=none,style=numeric-comp,firstinits=true]{biblatex} %style=authoryear-comp
\usepackage{url} % not crucial - just used below for the URL 

\expandafter\let\csname equation*\endcsname\relax
\expandafter\let\csname endequation*\endcsname\relax
\usepackage{amsfonts,amssymb,bm,amsthm}
\usepackage{ulem} %for strike through
\usepackage{array,booktabs}
\newcolumntype{C}[1]{>{\centering\let\newline\\\arraybackslash\hspace{0pt}}p{#1}}
\newcolumntype{R}[1]{>{\raggedleft\let\newline\\\arraybackslash\hspace{0pt}}p{#1}}
\usepackage[table]{xcolor}
\usepackage[pdftex]{hyperref}
\definecolor{darkblue}{rgb}{0,0,.75}
\hypersetup{colorlinks=true, breaklinks=true,
linkcolor=darkblue, menucolor=darkblue, 
urlcolor=darkblue,citecolor=darkblue}

\def\Title {Hypothesis-based acceptance sampling for modules F and F1 of the European Measuring Instruments Directive}

\newcommand{\Pac}{P_\text{ac}} 
\newcommand{\new}[1]{#1}%{{\color{red}{#1}}}
\newcommand{\SD}[1]{{\color{black}{#1}}}

\begin{document}

\title{\bf \Title}
\author{Katy Klauenberg$^1$\thanks{Corresponding author: Katy.Klauenberg@Ptb.de}, Cord A.\ M\"uller$^2$ and Clemens Elster$^1$\\
\small$^1$Physikalisch-Technische Bundesanstalt, Abbestr. 2-12, 10587 Berlin, Germany\\
\small$^2$Deutsche Akademie f\"ur Metrologie, Bayerisches Landesamt f\"ur Ma\ss{} und Gewicht,\\\small Wittelsbacherstr.~14, 83435 Bad Reichenhall, Germany}
\date{}
 \maketitle
	
\begin{abstract} 
 Millions of measuring instruments are verified each year before being placed on the markets worldwide. 
 In the EU, such initial conformity assessments are regulated by the Measuring Instruments Directive (MID). The MID modules F and F1 on product verification allow for statistical acceptance sampling,  whereby only random subsets of 
instruments need to be inspected. 
 This paper re-interprets the acceptance sampling conditions formulated by the MID. The new interpretation is contrasted with the one advanced in WELMEC guide 8.10, and three advantages have become apparent. 
 Firstly, an economic advantage of the new interpretation is a producers' risk bounded from above, such that measuring instruments with sufficient quality are accepted with a guaranteed probability of no less than 95\%. 
 Secondly, a conceptual advantage is that the new MID interpretation fits into the well-known, formal framework of statistical hypothesis testing. Thirdly, the new interpretation applies unambiguously to finite-sized lots, even very small ones.
 We conclude that the new interpretation is to be preferred and suggest re-formulating the statistical sampling conditions in the MID.  
 Re-interpreting the MID conditions implies that currently available sampling plans are either not admissible or not optimal. We derive a new acceptance sampling scheme and recommend its application.
\end{abstract} 


\noindent{Supplemental spreadsheet data is provided as an ancillary file, and further material for this article is available online.}

%\newpage
%\spacingset{1.45}
%------------------------------------------------------------------------- 
\section{Introduction}
\label{sec:intro} 
%-------------------------------------------------------------------------

 Statistical sampling is used in numerous fields, such as industry, medicine, election forecasts and many more, to acquire knowledge about an entire population by observing only a subset thereof. In legal metrology, sampling plans are applied, e.g., to assess the conformity of measuring instruments to be placed on the market \cite{MID}, to verify the quantity of product in prepackages \cite{OIMLR87}, and to re-verify  
utility meters at predefined periodic intervals \cite{OIMLG20,KlaEl18b}.

\begin{figure}
 \center
 \includegraphics[scale=1,clip,trim=0 10 0 30]{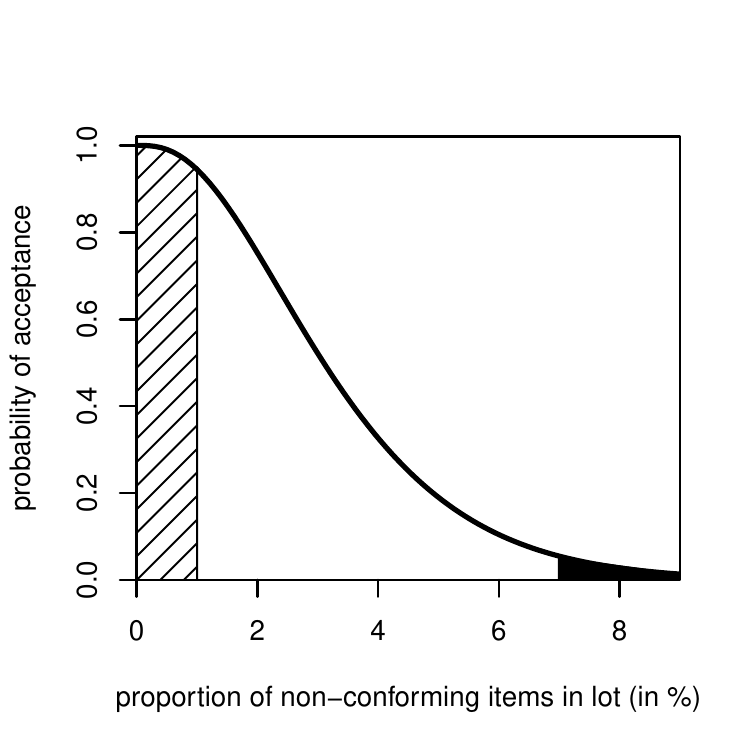}
 \caption{Operating characteristic (OC) curve for the sampling plan $(n,c)=(86,2)$ for a very large lot. Lots with less than 1\% non-conforming items are likely to be accepted (striped area), whereas lots with more than 7\% non-conforming items are likely to be rejected (shaded area).
  }
	\label{fig1}
\end{figure}

 Operating characteristic (OC) curves
 are a common tool for judging the performance of sampling plans, see e.g.\ \cite{FerGru46,Dun74,Dodg98,Mon09,Schil17}. They graphically display the probability of accepting a lot as a function of its proportion of non-conforming items. Figure~\ref{fig1} exemplifies such an OC curve for the acceptance sampling plan $(n,c)=(86,2)$, where $n=86$ items are sampled and the lot is accepted when at most $c=2$ non-conforming items are found. 
 Producers can easily infer from the OC curve which quality levels of their product will ensure a high probability of acceptance. For example, under the plan $(86,2)$, lots with less than 1\% non-conforming items are accepted in more than 95\% of the cases on average (figure~\ref{fig1}, striped area). In addition, consumers can infer which quality levels are likely to be rejected. Under the plan $(86,2)$, lots with more than 7\% non-conforming items are rejected in more than 95\% of the cases (figure~\ref{fig1}, black area).

%-------------
%-----------

 The steepness of its OC curve shows how well a sampling plan discriminates between good and bad quality. Therefore, sampling plans are often designed such that their OC curves meet certain conditions (e.g.\ \cite[part 2]{Dun74}, \cite[chap.\ 5]{Schil17}, \cite{ISO2859-1,ISO2859-2}). Prescribing two points of the OC curve determines a sampling plan when deciding between two attributes (e.g.\ conformance and non-conformance).
 
 A prominent example of such a two-point attribute sampling design in legal metrology is the European Measuring Instruments Directive (MID) \cite{MID}, which shall be the focus of this research. Since 2006 it has been harmonizing the market entry requirements for different types of measuring instruments, encompassing utility meters for water, gas, electricity and heat, as well as automatic weighing instruments, various material measures, taximeters, exhaust gas analyzers and many others. The MID's conformity assessment modules F and F1, for product verification by a notified body, provide the option to test either every instrument or to proceed by ``statistical verification''. For the latter, modules F and F1 require attribute sampling plans to ensure
\begin{quote}\label{QMID}
    \begin{itemize}
      \item[``(a)] a level of quality corresponding to a probability of acceptance of 95\%, with a non-conformity of less than 1\%;
      \item[ (b)] a limit quality corresponding to a probability of acceptance of 5\%, with a non-conformity of less than 7\%.'' 
    \end{itemize}
\end{quote} 
The wording of these two conditions is imprecise and thus requires a mathematical  interpretation in order to design sampling plans to be used in practice.

\begin{figure}
 \center
 \includegraphics[scale=1,clip,trim=0 10 0 30]{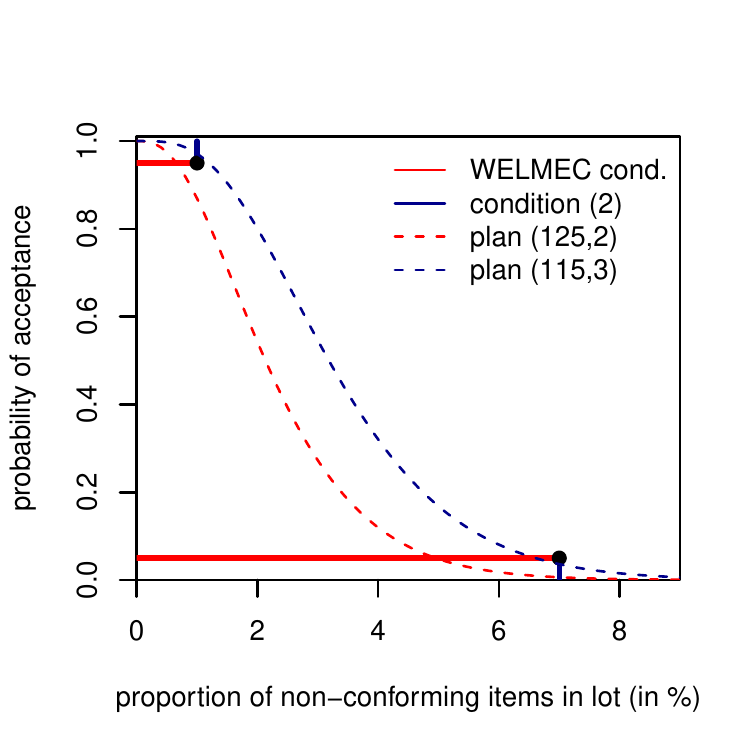}
	\caption{The WELMEC condition~\eqref{QWelm} and the new condition~\eqref{QOur} (thick red and blue line, respectively), and an example sampling plan for each (dashed lines). Both \eqref{QWelm} and \eqref{QOur} are interpretations of the MID conditions (a) and (b) in modules F and F1.}
	\label{fig2}
\end{figure}

In 2011 and 2018 the European Cooperation in Legal Metrology issued WELMEC guide 8.10 \cite{Welmec8.10} to generate sampling plans for \SD{Annexes} F and F1 of the MID, which proposes interpreting the conditions~(a) and~(b) as follows: 
\begin{equation}\label{QWelm}
 \text{``The OC curves have to be on the left hand side of the points mentioned'',}
\end{equation} 
referring to the points (1\%, 95\%) and (7\%, 5\%) displayed in figure~\ref{fig2} (black dots). Subsequently, \eqref{QWelm} will be called the WELMEC condition. Any OC curve passing through the thick red lines in figure~\ref{fig2} fulfills the WELMEC condition (such as the dashed red curve). The WELMEC guide \cite{Welmec8.10} provides sampling plans fulfilling this condition that are extracted from the well-known standard \cite{ISO2859-1}. This standard, however, does not cover the MID conditions to the full extent, as noted already in \cite{Welmec8.10}, because it addresses series of lots and different points of the OC curve. 
Under the same interpretation~\eqref{QWelm}, optimized sampling plans were recently derived in \cite{Muel19}. There, the sampling scheme proposed for practical use compromises between small sample sizes, small distances of the OC curve to the points (1\%, 95\%) and (7\%, 5\%) and the simplicity of the whole scheme.

This research proposes an alternative interpretation of the MID conditions~(a) and~(b): 
\begin{eqnarray}\label{QOur}
 &&\text{``The OC curves of sampling plans have to pass above or through}\\[-0.5ex]\nonumber
 &&\text{ the point (1\%, 95\%), and below or through the point (7\%, 5\%).''}
\end{eqnarray}
That is, any curve passing through the thick blue lines in figure~\ref{fig2} fulfills this condition (such as the dashed blue curve). Note especially the qualitative difference to the WELMEC condition~\eqref{QWelm} regarding the left point (1\%, 95\%). This alternative interpretation~\eqref{QOur} for (near to) infinite lot sizes is mainly motivated by its equivalence  to the hypothesis test
\begin{equation}\label{EqHypTest}
H_0\!: p \leq 1\%, \quad H_A\!: p\geq 7\%, \quad 
\text{with type I and II error rates } \alpha, \beta \leq 5\%\,,
\end{equation} as will be explained in section~\ref{sec:Hypo}. Indeed, the statistical framework of hypothesis testing is well established for designing sampling plans and has clear advantages. 
It bounds the type I and II error rates from \textit{above}, which in the present context symmetrically limits the risks of false decisions for both producers and consumers. This is in contrast to the WELMEC condition~\eqref{QWelm}, which  counterintuitively imposes a \textit{lower} bound on the producers' risk.  \new{That is, the WELMEC condition requires \textit{at least} 5\% rejection with 1\% non-conforming items.}
Also, the full inspection of lots is entirely compatible with hypothesis testing, 
whereas it does not generally satisfy the WELMEC condition, which is in fact ill-defined for finite lot sizes and does not cover very small lots (see section~\ref{subs:PotHypo} below).

Interpretation~\eqref{QOur} has, to our knowledge, not been proposed before. 
The following section~\ref{sec:Hypo} continues to argue that the hypothesis-based interpretation~\eqref{EqHypTest} of the MID conditions is more apt for product verification in legal metrology than the WELMEC interpretation~\eqref{QWelm}.
The authors are not aware of any standard for or guide to a sampling scheme that realizes test~\eqref{EqHypTest}. 
Section~\ref{sec:Plans} derives such a set of sampling plans for finite and infinite lot sizes and refers to the ancillary spreadsheet for a complete list. Subsequently, these plans are compared to the ones suggested in \cite{Muel19} for the WELMEC condition~\eqref{QWelm}. In order to ease the application of sampling according to the new interpretation~\eqref{EqHypTest} of the MID conditions~(a) and~(b), section~\ref{sec:Pract} derives a simplified sampling scheme which is a trade-off between the minimal sample size and a small number of different plans over all finite lot sizes.

Furthermore, false decisions about the conformity of measuring instruments can have vastly different consequences depending on the economic value or potential damage involved. 
Section~\ref{sec:Disc} will generally discuss more flexible alternatives to the MID's choice of fixing a single predefined producers' risk and a single predefined consumers' risk for all regulated measuring instruments. At present, the MID prescribes acceptance sampling by attribute, such that alternative approaches are beyond the scope here. However, more efficient sampling plans can be obtained, for example, by distribution-based attribute sampling \cite{ElKla20,KlaEl18}, variable sampling \cite{ISO3951-1}, series of lots \cite{ISO2859-1}, sequential sampling, by using prior knowledge \cite{ChalVer95,OHaSteCa05}, applying rectifying inspection \cite{Dun74} or statistical process control \cite{ISO11462-2,ISO7870}.

We conclude in section~\ref{sec:Concl} by proposing an unambiguous reformulation of the MID conditions~(a) and~(b) and by recommending an appropriate sampling scheme.

%------------------------------------------------------------------------- 
\section{Methods -- Hypothesis testing} 
\label{sec:Hypo}
%-------------------------------------------------------------------------
\subsection{Brief introduction}
\label{subs:IntroHypo}
 Statistical hypothesis testing is a formal framework to evaluate claims or statements on the basis of limited observations, e.g., to advance data-based knowledge in the sciences, for decision-making in criminal courts or for quality management in industry. Many, very different introductions to hypothesis testing exist and we refer the reader to, e.g., \cite{LehRo06,BliMu11,TaeKu14,Nuz14,Per15,Was16,KlaEl19}. Likewise, we refer to \cite{BenBer18,Was16,GreSe16,Ber03,HuBa03} for recent debates on hypothesis tests. 
 Subsequently, we present a very brief introduction, tailored to MID acceptance sampling, on how to specify hypothesis tests and derive sampling plans compatible with these. 

 The quantity of interest for MID conformity assessment in modules F and F1 is the proportion of non-conforming items $p$ in a predefined lot. Inspecting a random sample from the lot provides information on the proportion $p$, for instance to decide whether a specific value or range of $p$ is supported. This decision is based on the number of non-conforming items detected in the sample. 
In the long term, the decisions are more often correct when more items are sampled, while smaller samples are more economical. 
The decision rule for lot acceptance as well as the sample size required for a confident decision are thus of particular interest.

 A hypothesis test is formalized by stating two complementary hypotheses, $H_0$ and $H_A$. 
 The null hypothesis $H_0$ is typically chosen to be the proposition that can only be rejected by sufficient evidence to the contrary. Note that generally, samples can possibly disprove $H_0$, but cannot prove it \cite{GreSe16}. Say, we hypothesize that the lot is of good quality, i.e.\ it contains a proportion $p$ of non-conforming items lower than a  specified value. 
The alternative hypothesis $H_A$ contains the violations of the null hypothesis that shall be detected, namely a non-conforming proportion $p$ higher than another specified value.  
The hypotheses that match interpretation~\eqref{QOur} are
\begin{equation}\label{EqHyp}
 H_0\!: p \leq 1\%, \quad H_A\!: p\geq 7\%\,
\end{equation}
(compare e.g.\ \cite[chap.\ 8.1]{LehRo06} or  \cite[chap.\ 25.6]{Dun74} for hypothesis tests with indifference zone).

 The quality of a hypothesis test can be quantified by its type I and type II error rates. 
A type I error occurs when $H_0$ is rejected on the basis of the sample data, although $H_0$ is really true. In the present context,
 the producers of measuring instruments have an interest in avoiding this type of error; that is  avoiding the rejection of a lot due to finding a low-quality sample, although the lot is actually of sufficient quality $p\leq 1\%$.  
 The type I error rate $\alpha$ is thus also called the producers' risk here. 
 A type II error occurs if $H_0$ is not rejected, although $H_A$ is true, i.e.\ when a lot is accepted due to a good-quality sample although the lot is really of bad quality $p \geq 7\%$. 
 End users of measuring instruments have an interest in avoiding this error, such that the type II error rate $\beta$ is also called the consumers' risk. Given the hypotheses~\eqref{EqHyp}, our MID interpretation~\eqref{QOur} for infinite lot size is equivalent to the risk bounds
\begin{equation}\label{EqRisks}
 \alpha  \leq 5\% \text{ and } \beta \leq 5\% \,.
\end{equation}
The hypotheses~\eqref{EqHyp} and risks~\eqref{EqRisks} uniquely determine a hypothesis test. They are equivalent to formulation~\eqref{EqHypTest}, as well as to the interchanged hypothesis test $H_0\!: p\geq 7\%, \  H_A\!: p \leq 1\%$ as long as the risks are equal, $\alpha=\beta$.
 
 For testing proportions, the number of non-conforming items found in a random sample is the natural test statistic (c.f.\ \cite[Ex 3.4.1-2]{LehRo06}); its value determines whether the lot is accepted or not. 
 The test statistic, say $x$, follows a binomial distribution when the $n$ items of a sample are independent (e.g.\ drawn from an infinite lot $N=\infty$ or with replacement) such that each item has the same probability $p$ to be non-conforming,
\begin{equation}\label{EqBinom}
P(x;n,p,N) = \binom{n}{x}p^x(1-p)^{n-x}\,,
\end{equation}
and it follows a hypergeometric distribution when the sample is drawn without replacement from a lot of finite size $N$ containing $pN\in \mathbb{N}$ non-conforming items,
\begin{equation}\label{EqHyper}
P(x;n,p,N)= \frac{\binom{pN}{x}\binom{N-pN}{n-x}}{\binom{N}{n}} \,.
\end{equation}

Given a sample size $n>0$, 
there is a maximum number $c_n$ of non-conforming items such  that the risks~\eqref{EqRisks} are not exceeded.
The decision rule of the test then states that for $x\leq c_n$ hypothesis $H_0$ cannot be rejected, and the lot is accepted. 
The acceptance probability
is the cumulative distribution
\begin{equation}\label{Pacc}
\Pac(c_n;n,p,N) = \sum_{x \leq c_n}P(x;n,p,N)
\end{equation}
with density function~\eqref{EqBinom} or~\eqref{EqHyper} for infinite or finite lot sizes, respectively. 

Under the hypothesis test~\eqref{EqHypTest}, many sampling plans  $(n,c_n)$ are admissible. 
We define that the optimal sampling plan is the pair $(n^*\!,c^*)$ with minimal sample size, i.e.\ 
$n^*\!\leq n$ for all admissible plans and $c^*=c_{n^*}$.% 
\footnote{Depending on context and perspective, optimality may be defined differently, for instance, by minimizing a certain risk rather than sample size. More generally, one can combine several parameters in a cost function to be minimized, potentially resulting in very different sampling plans.}
For this optimal sampling plan, the acceptance probability as a function of the non-conforming proportion $p$ and lot size $N$ shall be denoted by $\Pac^*(p,N):=\Pac(c^*;n^*,p,N)$.  
Under the binomial distribution~\eqref{EqBinom}, this optimal sampling plan turns out to be  $(n^*\!,c^*)=(109,3)$ with risks 
 $\alpha=1-\Pac^*(0.01,\infty)=2.43\%$ and $\beta=\Pac^*(0.07,\infty)=4.85\%$; 
see figure~\ref{fig4} for the corresponding OC curve.
The more general case, namely optimal sampling plans for finite lots under the hypergeometric distribution~\eqref{EqHyper}, 
is derived in section~\ref{sec:Plans}. 
A complete list of optimal sampling plans for all finite lot sizes can be found in the ancillary spreadsheet, whereas a reduced scheme of near-optimal sampling plans, which is compact and easier to handle, is proposed in section~\ref{sec:Pract}.

 Let us complete this introduction with a conceptual comparison of our  hypothesis-based MID interpretation~\eqref{EqHypTest} with the WELMEC interpretation~\eqref{QWelm}.

\subsection{Conceptual comparison of MID interpretations}
\label{subs:PotHypo}

\begin{figure}
 \center
 \includegraphics[scale=1,clip,trim=0 10 0 30]{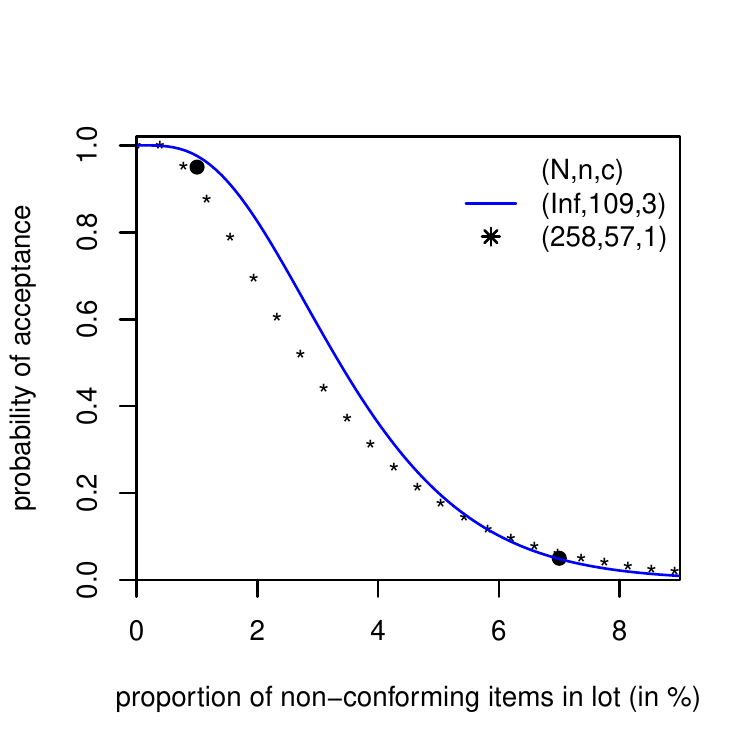}
	\caption{Operating characteristic (OC) of two optimal sampling plans---plan $(109,3)$ for infinite lot size and plan $(57,1)$ for %\new{\sout{finite}} 
	lot size $N=258$%\new{\sout{, respectively}}
	---as generated by the acceptance probability~\eqref{Pacc} for the hypothesis test~\eqref{EqHyp}.}
	\label{fig4}
\end{figure}

First, let us discuss the (idealized) case of infinite lot size, for which the OC curve is continuous and runs through the two points $(1\%,1-\alpha)$ and $(7\%,\beta)$ for some values $\alpha$ and $\beta$. Our interpretation~\eqref{QOur} and hypothesis test~\eqref{EqHypTest} (or~\eqref{EqHyp} and~\eqref{EqRisks}) are then equivalent%
\footnote{For infinite lot size, the OC is continuous such that the points $(1\%,1-\alpha)$ and $(7\%,\beta)$ exist. Due to the monotonicity of the OC curve, sampling plans fulfilling condition~\eqref{QOur} thus also fulfill test~\eqref{EqHyp}-\eqref{EqRisks} and vice versa. Compare also \cite[7.2.3 and 25.6]{Dun74}.}, such that interpretation~\eqref{QOur} inherits all properties from the sound framework of statistical hypothesis testing. In particular, linking the MID requirements~(a) and~(b) to a single statistical hypothesis test makes the risks transparent for all parties concerned. Hypothesis test~\eqref{EqHypTest} symmetrically limits the producers' risk of rejecting lots of good quality, as well as the consumers' risk of accepting lots of bad quality. 
 Both risks are bounded in advance---at 5\% in the MID setting as expressed by our interpretation~\eqref{QOur}. 

 In contrast, the WELMEC interpretation~\eqref{QWelm} cannot be formulated as a single hypothesis test. Only in hindsight, may the sampling plans resulting from~\eqref{QWelm} be linked to several hypothesis tests with different properties. Specifically, one finds quite different risks for producers and consumers. 
Moreover, valid sampling plans can yield arbitrarily low acceptance probabilities for good quality levels 
 and thus unbounded type I error rates. 
 This was noted in \cite{Muel19}, where optimized sampling plans with producers' risks of up to 35\% are encountered.   
 Indeed, the WELMEC interpretation~\eqref{QWelm} does not fix an upper bound for the producers' risk at 1\% quality, but a lower bound, which is unusual in quality control \cite{Schil17}. 
 The requirement that the OC curve passes to the left of the point  $(1\%,1-\alpha)$ 
 could rather be viewed as setting an additional bound for the consumers' risk.  
 That is, consumers are protected at two quality levels, whereby lots with more than 1\% and more than 7\% non-conforming items will be wrongly accepted in less than 95\% and less than 5\% of cases, respectively---but producers enjoy no protection whatsoever against type I errors. 
 Clearly, the producers' interest  
 is entirely disregarded when interpreting the MID conditions according to the WELMEC guide.

Second, we turn to the (practically most relevant) case of finite lot sizes. 
For lots of size $N$ containing $pN\in\mathbb{N}$ non-conforming items, a sampling plan's OC is a set of discrete points. 
In that case, the WELMEC condition~\eqref{QWelm} is ambiguous, because what is meant by requiring a ``curve'' consisting of discrete points to be on the left-hand side of two given points? 
The OC points will almost never lie exactly on the probability levels of 95\% and 5\% (nor will they be located at the quality levels $p=0.01$ and $p=0.07$ unless $N$ is a multiple of 100), so that there is no means of deciding whether the ``left-hand side" criterion is fulfilled or not.  
For example, it is unclear whether the sampling plan $(57,1)$ for $N=258$, as displayed in figure~\ref{fig4} by stars, is admissible. 
Hence, the WELMEC interpretation~\eqref{QWelm} is generally ill-defined for finite lot sizes, and an additional clarification is required. 
In \cite{Muel19} a continuous interpolation is proposed, but alternatives are conceivable, such as constraining all discrete points $p\geq 1\%$ and $p\geq 7\%$ of the OC graph to lie below the 95\% and 5\% levels, respectively, which was considered in \cite[App.]{Muel19}.  
The sampling plan $(57,1)$ would be admissible for the latter, pointwise criterion, but not for the former, continuous criterion with a consumers' risk of more than 5\% at the (fictitious) quality level $p=7\%$. 
Certainly, also the modified interpretation~\eqref{QOur} based on the notion of a continuous OC curve suffers from a similar ill-definedness.  
In contrast, the hypothesis test~\eqref{EqHypTest} is well-defined for infinite as well as all finite lot sizes.
For example, for $N=258$, the optimal sampling plan $(57,1)$ has type I and II error rates of at most 4.8\% and 4.9\% for all quality levels $p\leq 1\%$ and $p\geq 7\%$, respectively 
(further details can be found in section~\ref{sec:OptHypTest}). 
Additionally, the WELMEC interpretation~\eqref{QWelm} disregards the producers' interest, whereas the hypothesis test \eqref{EqHypTest} sets an upper bound for both the producers' and consumers' risks, for any lot size, as explained above.

There are thus three advantages to basing MID acceptance sampling on hypothesis tests: firstly, the theoretically sound framework of statistical hypothesis testing, secondly, the limited risk for producers and thirdly, no ambiguity for finite lot sizes.

%------------------------------------------------------------------------- 
\section{Results -- Sampling plans} 
\label{sec:Plans}
%-------------------------------------------------------------------------

Let us now derive the optimal sampling plans for the hypothesis test~\eqref{EqHypTest} for all lot sizes. 
These plans, whose calculation is explained in the following section~\ref{sec:OptHypTest}, are compared to the optimal sampling plans for the WELMEC condition~\eqref{QWelm} in section~\ref{sec:complans}. 

%-------------------------------
\subsection{Optimal sampling plans for the MID hypothesis test}
\label{sec:OptHypTest}

\begin{figure}
 \hspace{-0.5em}
 \includegraphics[width=\textwidth,clip,trim=3 10 10 40]{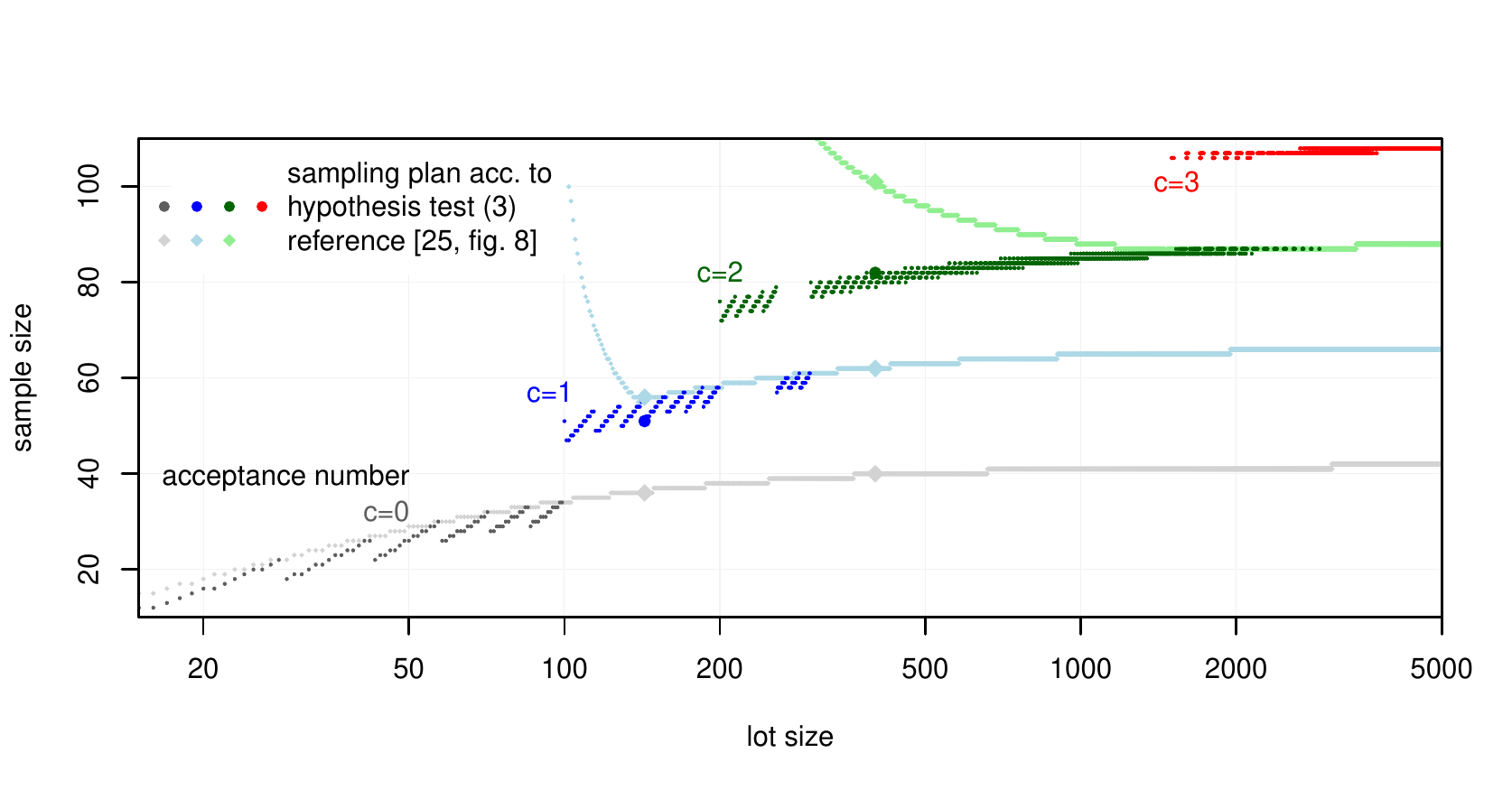}
	\caption{Minimal sample size for hypothesis-based MID sampling plans (dark dots) and for WELMEC-based sampling plans as in \cite[fig.\ 8]{Muel19} (light diamonds). Colors represent the corresponding acceptance number and larger points the plans discussed in the text.}
	\label{fig3}
\end{figure}

As explained in section~\ref{subs:IntroHypo}, the optimal sampling plan $(n^*,c^*)$ under the hypothesis test~\eqref{EqHypTest} is found by looking for the smallest sample size $n^*$ and the corresponding acceptance number $c^*=c_{n^*}$ such that the type I and II error rates do not exceed 5\%. Because the acceptance probability~\eqref{Pacc} is a decreasing function of $p$, it suffices to check the value that is closest to and not above 1\%, and the value closest to and not below 7\%, i.e.\     
$\new{p_\alpha} := \lfloor 0.01N\rfloor/N$ and $\new{p_\beta} := \lceil 0.07N\rceil/N$, respectively.  
We identify the producers' risk (type I error rate) 
\begin{equation}\label{Defalpha}
\alpha = 1 - \Pac(c;n,\new{p_\alpha},N)
\end{equation} 
and the consumers' risk (type II error rate) 
\begin{equation}\label{Defbeta}
\beta = \Pac(c;n,\new{p_\beta},N),   
\end{equation} 
knowing that all smaller (larger) quality levels $p\leq \new{p_\alpha}$ ($p\geq \new{p_\beta}$) cannot have larger risks, due to the monotonicity of~\eqref{Pacc}. 
Then, according to criterion~\eqref{EqRisks}, a sampling plan $(n,c)$ is admissible if both $\alpha,\beta\leq 0.05$, which is simple to check. 

For infinite lot size, where $\new{p_\alpha}=0.01$ and $\new{p_\beta}=0.07$ are points on the continuous OC curve, the plan $(n^*\!,c^*)=(109,3)$ is optimal. 
The OC curve of this plan is shown in figure~\ref{fig4} as the continuous line. 
When the lot size decreases, there is a global tendency for smaller samples to become admissible because the limiting binomial distribution~\eqref{EqBinom} conservatively approximates the probability to draw non-conforming items from the lot, which actually follows the hypergeometric distribution~\eqref{EqHyper}, cf.\ \cite[chap.~4]{Schil17} and \cite[sec.~3.1]{Muel19}.  For example, for $N=258$, the sampling plan $(n^*\!,c)=(57,1)$ is the one with the smallest sample size fulfilling the hypothesis test~\eqref{EqHypTest} and shown in figure~\ref{fig4} by symbols. 

The optimal sampling plans for all finite lot sizes $N\leq 10^4$ are computed by systematically checking conditions~\eqref{Defalpha} and~\eqref{Defbeta}. These plans are listed in the ancillary spreadsheet and most are
displayed in figure~\ref{fig3} as dark \new{dots}, with colors indicating the different acceptance numbers. 
An R Shiny app \cite{Shiny} calculates the optimal MID sampling plan for a given lot size \cite{ourRShinyapp}.
The smallest sample size $n^*$ is not an increasing function of the lot size $N$, but appears as a seesaw pattern in figure~\ref{fig3}, due to the inherent discreteness of the hypergeometric distribution,  
cf.\ \cite[App.]{Muel19}. 
The simplified sampling scheme derived in section~\ref{sec:Pract} addresses this issue. 

\subsection{Comparison of sampling plans and their properties}
\label{sec:complans}
Let us compare the sampling plans for the hypothesis test~\eqref{EqHypTest} with the ones according to the WELMEC condition~\eqref{QWelm}. 
In particular, we will compare our optimal plans and their properties with the ones in \cite[fig.\ 8]{Muel19}. The latter plans are reproduced in figure~\ref{fig3} (light colored diamonds) and show the minimal sample size for acceptance numbers $c=0,1,2$. 

At the outset, we remark that for any fixed sampling plan, the producers' and consumers' risks of the hypothesis-test interpretation will be
smaller than or equal to the risks of the WELMEC interpretation as used in \cite{Muel19}. 
Indeed, due to the continuous interpolation, the risks cited in \cite{Muel19}, say $\alpha_\text{cont},\beta_\text{cont}$, are always evaluated at the quality levels $0.01$ and $0.07$, which may not be realized in the lot. 
Hypothesis tests calculate the risks $\alpha, \beta$ for the realized levels $\new{p_\alpha} =\lfloor 0.01N \rfloor/N \leq 0.01$ and 
$\new{p_\beta}=\lceil 0.07N \rceil/N \geq 0.07$. 
Especially when $\new{p_\alpha} \ll 0.01$ ($\new{p_\beta}\gg 0.07$), the resulting difference between the risks $\alpha\leq \alpha_\text{cont}$ ($\beta\leq \beta_\text{cont}$) can be quite pronounced. 
Notably, for lot sizes $N < 100(c+1)$, the producers' risk $\alpha$ is identically zero because a lot with quality $p \leq 1\%$ contains 
$pN \leq N/100 < c+1$ 
defective items and is thus never rejected.

We now compare the sampling plans proceeding from smaller to larger lot sizes.
  
For very small lots of size $N<15$, hypothesis test~\eqref{EqHypTest} requires a full inspection $n=N$. In contrast, under the WELMEC condition~\eqref{QWelm}, lots with $N<11$ cannot be inspected at all, because neither a full inspection nor any other sampling plan is admissible.

It is yet another, logically satisfying advantage of the hypothesis-based interpretation that a full inspection is admissible for all $N$.

For lot sizes $15 \leq N < 100$, both MID interpretations lead to sampling plans with acceptance number $c=0$. Due to the risk difference as explained at the outset, the sample sizes in \cite{Muel19} are larger than or equal to ours. For example, for $N=43$ our sampling plan is $(22,0)$ with risks $\alpha=0$, $\beta = 0.048$, while the plan in \cite{Muel19} is $(27,0)$ with $\alpha_\text{cont} = 0.343,\, \beta_\text{cont} = 0.045$. Retrospectively interpreting the sampling plan $(27,0)$ as a hypothesis test gives risks $\alpha=0,\, \beta \leq 0.015$.

The lot sizes $N=100,101$ lead to counterintuitive results in \cite{Muel19}, as a consequence of the counterintuitive lower bound on the producers' risk in the WELMEC interpretation.   
In particular, one would expect that a full inspection of the lot with $c=1$ observed non-conforming item would be an admissible sampling plan under the MID, because one can conclude that $p=1/N\leq 0.01$ with certainty. However, such a fully inspecting sampling plan $(N,1)$ is not admissible under the WELMEC condition~\eqref{QWelm}, because the probability of acceptance is $1-\alpha=1$ and $1-\alpha = 0.959$, respectively (due to the analytic continuation for the latter) and thus exceeds 95\%. Likewise and also against intuition, the sampling plans $(N,c)$ for lot sizes $N=100c$ and $N=100c+1$ are not admissible. 

For lot sizes $N > 101$, \cite{Muel19} does not specify a unique sampling plan, but offers minimized sample sizes for two or more acceptance numbers, 
between which the end users have to choose. 
In contrast, optimal sampling plans for hypothesis tests are unique (as defined in section~\ref{subs:IntroHypo}, the sample size is minimized globally). 
Larger sample sizes as well as sampling plans with larger acceptance numbers (and suitable larger sample sizes) are admissible as well. 
Obviously, the sampling plans in \cite{Muel19} with $c=0$ have smaller sample sizes than our plans with $c=1$. However, the producers' risk of rejecting good quality lots is much larger for plans with $c=0$ than for $c\geq 1$. (The consumers' risks are usually similar, often just below 0.05 and will not be discussed further.)
Let us discuss two examples: the ones highlighted in figure~\ref{fig3}. First, for $N=143$ the plan $(36,0)$ is admissible in \cite{Muel19} and has a risk of $\alpha = 0.252$ ($\alpha_\text{cont} = 0.340$). That is, more than a quarter of all lots containing one defective item (i.e.\ quality level $p=1/143 < 1\%$) will be 
rejected under this scheme. 
In contrast, the plan $(51,1)$ is admissible under test~\eqref{EqHypTest} and has a risk of $\alpha=0$. Also admissible in \cite{Muel19} is the plan $(56,1)$ for $N=143$, with a risk of $\alpha=0$ ($\alpha_\text{cont} = 0.055$). 
A second example is $N=400$. According to \cite{Muel19}, the plans $(40,0)$, $(62,1)$ and $(101,2)$ are optimal, with producers' risks $\alpha_\text{cont} \approx \alpha$ of 0.345, 0.115 and 0.051, respectively. That is, the WELMEC interpretation allows a zero-acceptance sampling plan, where more than a third of lots with a quality level $p=1\%$ are rejected.
In contrast, the optimal plan for hypothesis test~\eqref{EqHypTest} is $(82,2)$ with a risk of $\alpha = 0.028$.

A salient feature of the WELMEC-optimized sampling plans, apparent in figure~\ref{fig3}, is the sharp rise of the required sample size with acceptance numbers $c\geq 1$ for lot sizes decreasing below $100c$. This very awkward, if not outright pathological feature, is a consequence of the \textit{lower} bound on the producer's risk, which forces an ever sharper downturn of the OC graph in order to stay below the point $(1\%,95\%)$ of the MID condition (a). For the hypothesis-based interpretation, no such pathological behavior is encountered. Globally, sample size and acceptance number increase quite naturally with growing lot size. 

For larger lot sizes $1500\leq N < 2900$, our optimal plans for hypothesis test~\eqref{EqHypTest} irregularly alternate between $c=2$ and $c=3$, while for $N\geq 2900$ only $c=3$ provides optimal plans. 
Acceptance numbers $c\geq 3$ are generally admissible under the WELMEC condition~\eqref{QWelm} as well, but were not considered in \cite{Muel19} because of the larger required sample sizes and dissymmetric risks (cf.\ \cite[Sec.\ 2]{Muel19}). 
For (near to) infinite lot size, \cite{Muel19} derives the sampling plan $(88,2)$ with $\alpha_\text{cont}=\alpha = 0.0587$, while the smaller plans $(66,1)$ and $(42,0)$ remain possible as well, albeit with higher risks of 0.141 and 0.344, respectively. 
For our test~\eqref{EqHypTest}, the optimal plan is $(109,3)$ with $\alpha = 0.0236$. 
In comparison, the hypothesis test requires a larger sample size (roughly by factors of 1.25, 1.65 and 2.60, respectively). In return, the chance of rejecting lots of acceptable quality is reduced substantially (roughly by factors of 2.5, 6.0, and 15, respectively). 

To summarize, the sampling plans that \cite{Muel19} developed for the WELMEC condition~\eqref{QWelm} provide a choice between different acceptance numbers and also provide small sample sizes, at the price of an elevated producers' risk. The latter is, by construction, always larger than for the hypothesis test~\eqref{EqHypTest}. In particular, the smaller the acceptance number, the larger this risk, and rejecting a third of the lots with acceptable quality is not rare. 
A further disadvantage of the WELMEC plans are their counterintuitive plans 
with acceptance number $c$ for lot sizes $N \leq 100c+1$
as well as their inability to handle full inspections and very small lot sizes. For practical use, an undesirable feature of our plans is their very detailed step structure where, due to discretization effects, sample size and acceptance number can locally decrease with an increasing lot size. This issue is addressed in the following section.   

\section{Results -- Simplified sampling plans}\label{sec:Pract}

\begin{figure}
  \hspace{-0.5em}
 \includegraphics[width=\textwidth,clip,trim=3 10 10 40]{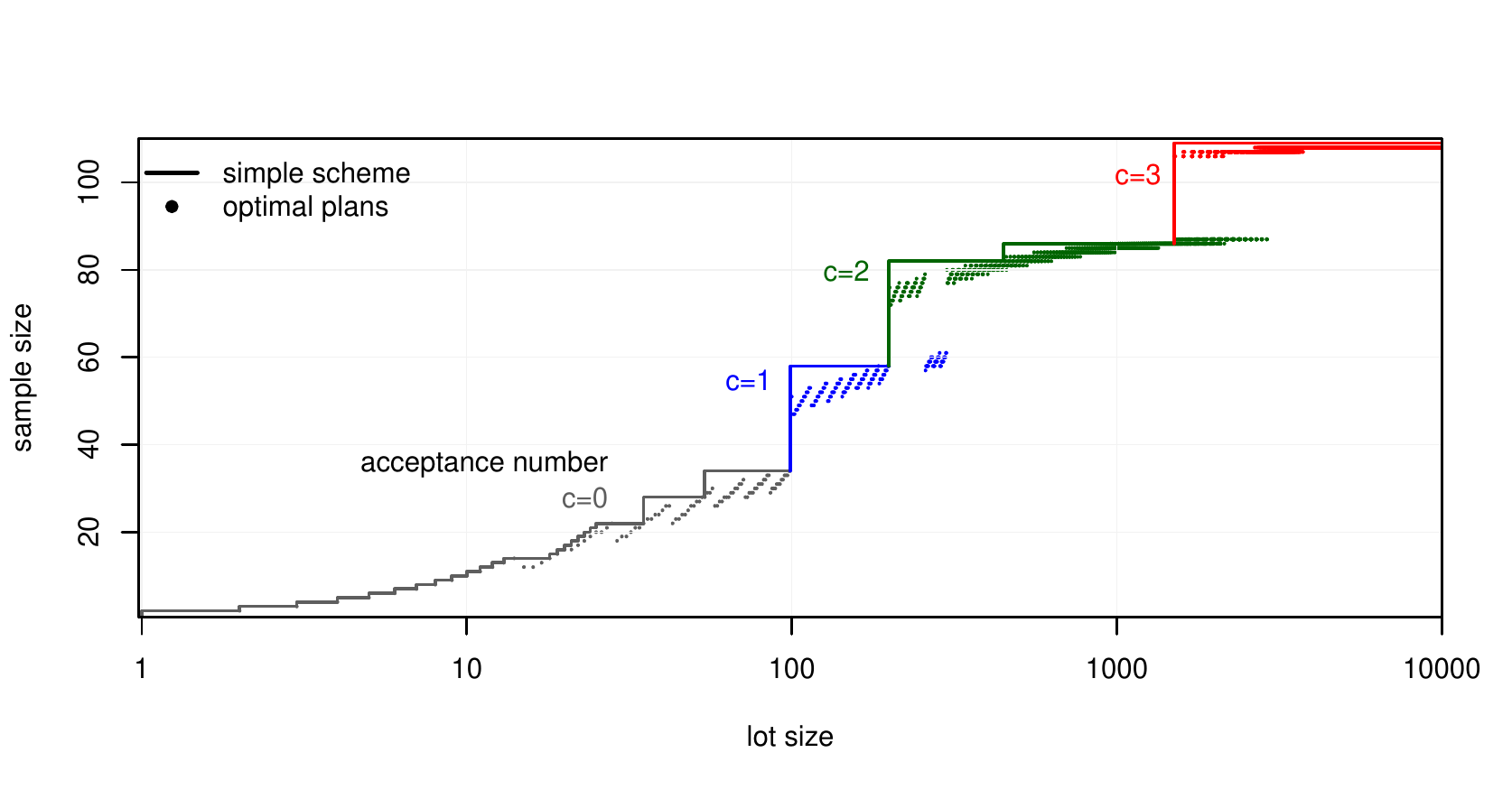}
	\caption{Optimal sampling plans (dots) together with a proposal for a simplified, nearly optimal sampling scheme \new{(solid line)} for hypothesis-based MID acceptance sampling as listed in table~\ref{tab_nmin_N_c1-3_simplified}. The colors distinguish different acceptance numbers.}
	\label{fig5}
\end{figure}

While the optimal sample size and the acceptance number for our plans globally increase with increasing lot sizes, these values may also decrease locally, due to the discretization of quality levels to $pN\in\mathbb{N}$. 
The resulting, quite detailed step structure is displayed in figure~\ref{fig3}. 
In order to arrive at a more practical and intuitively reasonable sampling plan, one may define larger intervals of lot sizes and, for each of these, propose the same admissible, albeit partially sub-optimal, sampling plan.
Such effective sampling schemes are not unique, and their ``construction is more art than it is science'' \cite[p.\ 210]{Dun74}. 
One possible proposal for a reasonably simplified, yet nearly optimal sampling plan is listed in table~\ref{tab_nmin_N_c1-3_simplified} and displayed by lines in figure~\ref{fig5}.

%-----------TABLE---
\begin{table}
\setlength{\tabcolsep}{0pt}
\rowcolors{3}{gray!25}{white}
\begin{center}
\begin{tabular}{*{2}{C{4em}}|*{2}{C{2.5em}}|*{2}{C{2cm}}|*{2}{C{2cm}}}
\multicolumn{2}{c|}{Lot size $N$}	& 
    \multicolumn{2}{c|}{Sample}& 	
    \multicolumn{2}{c|}{Producers' risk $\alpha$ [\%]} & 
    \multicolumn{2}{c}{Consumers' risk	$\beta $ [\%]}\\
from  & to 	&$n$	&	$c$ & from & to& from & to \\
\hline 
1 &  14 &   $N$ &   0   & 0  & 0  & 0  & 0 \\ 
%15 &  17/18 & {\small $\max(N-4,14)$} &   0   & 0  & 0  & 0  & 2.21 \\ 
15 &  18 & 14 &   0   & 0  & 0  & 0  & 3.92 \\ 
19 &  25 &  $N-4$ &   0   & 0  & 0  & 2.00  & 3.51 \\ 
26 &  35 &   22 &   0   & 0  & 0  & 0.96  & 4.37 \\ 
36 &  54 &  28 &   0   & 0  & 0  & 0.78  & 4.73 \\ 
55 & 99 &  34 &  0   & 0  & 0  & 0.93  & 4.68 \\ 
100 & 199 & 58 & 1 & 0  & 0  & 1.00  & 4.84 \\ 
200 & 449 & 82 & 2 & 0  & 2.85   & 1.97  & 4.96 \\ 
450 & 1499 & 86 & 2 & 1.74  & 4.98  & 3.36  & 4.99 \\ 
1500 & $\infty$ & 109 & 3 & 1.55  & 2.43  & 4.07  & 4.85 
\end{tabular}    
\end{center}
\caption{%
Proposal for a simplified, nearly optimal sampling scheme for hypothesis-based MID acceptance sampling. 
By construction, the producers' and consumers' risk never exceed 5\%. For lot sizes $N < 100(c+1)$, the producers' risk is zero because a lot with quality $p \leq 1\%$ contains 
$pN < c+1$ defective items and is thus never rejected.} 
\label{tab_nmin_N_c1-3_simplified}
\end{table}

The proposed scheme is compact, nearly optimal and guarantees both consumers' and producers' risks of below 5\%. 
In comparison, the simplified scheme in \cite{Muel19} 
requires \SD{choosing} between different acceptance numbers, and offers
 (slightly) smaller sample sizes. The hypothesis-based scheme 
guarantees substantially smaller producers' risks (especially for small acceptance numbers) and accommodates small lot sizes $N\leq 20$. 
Importantly, 
both simplified schemes yield sample sizes for large lots $N>1200$ that are substantially smaller than the ones extracted by the WELMEC guide \cite{Welmec8.10}  from the standard \cite{ISO2859-1}.

%------------------------------------------------------------------------- 
\section{Further discussion: adaptive test parameters } 
\label{sec:Disc}
%------------------------------------------------------------------------- 

The MID prescribes certain quality and risk levels for conformity assessment in  modules F and F1, with the implications discussed so far.   
However, the question arises of whether it is reasonable for the MID to fix these values in the first place, considering the aim of such a regulating directive.   
Is it sensible to provide a single numerical value for each risk level; and to fix numerical values at all ``so as to ensure a high level of protection of the aspects of public interest'' \cite{MID}?

The MID applies to a rather diverse set of measuring instruments, dedicated to  diverse measurement tasks. They comprise utility meters, automatic weighing instruments, taximeters, exhaust gas analyzers, measuring systems for quantities of liquids other than water, material measures and dimensional measuring instruments. These instruments are legally regulated for reasons of, among other things, public health and safety, 
the protection of the environment and consumers, and fair trade. 
When these instruments measure incorrectly, their errors cause various costs, lead to various kinds of damage, and generally have various implications for the public. Consequently, the same fraction (say, 7\%) of non-conforming, but  accepted devices will have very different consequences, and the acceptance probability for such devices ought to be adaptive. 

Future revisions of the European directive may thus consider variable levels of risks and quality, which can then be tailored to categories of measurement tasks. 
That is, consumers' risks substantially smaller than 5\% and/or quality levels substantially better than 7\% could be required for instruments potentially causing major damage, and more relaxed values could be tolerated for measurements where errors can only cause minor damage. 
For example, mis-measuring power meters in nuclear power plants can have more serious consequences than mis-measuring drinking cups in fast-food restaurants. An adaptive regulation could help to enforce public interest that is quantitatively comparable in all areas of legal metrology. 

Furthermore, the economic value of individual measuring instruments differs greatly. Thus, for producers discarding up to 5\% of measuring instruments due to type I errors of the sampling procedure, generates quite different costs. 
A more flexible regulation could provide an adaptive value for the tolerable producers' risk, which could be tuned according to production and verification costs or could potentially be set by the producers themselves. Each producer could then balance the expenses of sampling with the expenses of wrongly discarding products that actually conform to regulations. 
Indeed, we see no reason why the MID must restrain the market entrance in modules F and F1 in a highly inflexible manner, whereas the choice of statistical protection in other modules is left entirely to the producers.  

Greater flexibility for the producers as well as increased adaptivity to potential damage to public interest---both aims are within reach in the dawning age of digitalized legal metrology, where regulations need not be cast into the straitjacket of a single table with a few sampling plans. Instead, optimal plans could be computer-based and custom-tailored to variable parameters in future research, 
but should in any case be based upon the solid framework of statistical hypothesis testing.

%------------------------------------------------------------------------- 
\section{Conclusion} 
\label{sec:Concl}
%------------------------------------------------------------------------- 

The European Measuring Instruments Directive \cite{MID} formulates acceptance sampling conditions for statistical conformity assessment in modules F and F1 in an ambiguous manner, such that a mathematical interpretation is required.  
Under the interpretation of the WELMEC guide 8.10 \cite{Welmec8.10}, an optimized sampling scheme was proposed recently \cite{Muel19}. With the research presented here, we  highlight several shortcomings of the WELMEC interpretation and their consequences, develop a new interpretation based on hypothesis testing and track the ensuing changes down to optimal sampling plans.  

 The new MID interpretation proposed here is based on formal hypothesis testing, a well-known statistical framework. Therefore, this interpretation leads unambiguously to sampling plans for finite lot sizes, and it allows 100\% inspection of all lots---unlike the WELMEC interpretation. Notably, the new sampling plans symmetrically limit the risks for consumers and producers to the same value of 5\%, whereas under the WELMEC interpretation, producers' risks are unbounded and rise up to more than 30\% in realistic cases. Consumers' interests are preserved 
 at the previous level. 

We recommend a clarifying reformulation of the MID with only a minor change in wording.
All previously mentioned advantages are rendered by formulating no.~5.3 in module F and no.~6.4 in module F1 within Annex II of the MID as follows:
\begin{quote}\label{QMIDNew}
``The sampling system shall ensure:
 \begin{itemize}
  \item[(a)] a probability of acceptance of \textit{no less than} 95\% \textit{for} levels of quality of 1\% non-conformity \textit{and} less;
	\item[(b)] a probability of acceptance of \textit{no more than} 5\% \textit{for} a limit quality of 7\% non-conformity \textit{and more}.''
 \end{itemize}
\end{quote}
The modification eliminates the present deficiencies and, besides reordering the sentence, only slightly changes the current wording (the few new or modified words are highlighted).  
 Condition (a) now reflects the null hypothesis in test~\eqref{EqHypTest} with a type I error rate $\alpha\leq 5\%$, and condition (b) the alternative hypothesis with a type II error rate $\beta \leq 5\%$. 

 For the new formulation of the MID, we have developed optimal sampling plans and an easy-to-apply, well-behaved sampling scheme for a wide range of lot sizes. These sampling plans are generally different from the optimal plans for the WELMEC interpretation: Lots with less than 100 instruments require smaller sample sizes.  Larger lots require larger sample sizes. These extra expenses for the sampling procedure should be more than compensated for by reliably reducing the risk of unjustly discarding good-quality products---which is also an environmental advantage. 

Finally, as an outlook to future developments in legal metrology, we have pointed out the advantages of more flexible formulations, with variable parameters, when it comes to specifying criteria for statistical acceptance sampling. 

\printbibliography

\end{document}